# POLARIZABILITIES OF AN ANNULAR CUT AND COUPLING IMPEDANCES OF BUTTON-TYPE BEAM POSITION MONITORS

Sergey S. Kurennoy

Physics Department, University of Maryland, College Park, MD 20742, USA

Abstract

The longitudinal and transverse coupling impedances of a small discontinuity on the accelerator chamber wall can be expressed in terms of the electric and magnetic polarizabilities of the discontinuity. The polarizabilities are geometrical factors and can be found by solving a static (electric or magnetic) problem. However, they are known in the explicit analytical form only for a few simple-shaped discontinuities, for example, for an elliptic hole in a thin wall. In the present paper the polarizabilities of a ring-shaped cut in the wall are obtained. The results are applied to calculate the coupling impedances of button-type beam position monitors.

## I. Introduction

The coupling impedances of a small discontinuity on the wall of the vacuum chamber of an accelerator have been calculated in terms of the polarizabilities of the discontinuity [1], [2], [3]. The basic idea of the approach used is related with the Bethe theory of diffraction by small holes [4], which shows that fields produced by a hole can be approximated by those due to effective dipoles induced on the hole by an incident (beam) field. The magnitudes of the effective electric $P$ and magnetic $M$ dipoles are expressed through the incident fields $E_\nu^h, H_\tau^h$ at the hole location without hole [4], [5]

$$P_\nu = -\chi \varepsilon_0 E_\nu^h / 2; \quad M_\tau = \psi H_\tau^h / 2 , \tag{1}$$

where $\chi$ is the electric polarizability and $\psi$ is the magnetic susceptibility of the hole, $\hat{\nu}$ is the normal vector to the hole plane, and $\hat{\tau}$ is the tangential one. In general, $\psi$ is a symmetric 2D-tensor, but we will consider here only axisymmetric holes.

The hole polarizabilities are known in an analytical form only for a few simple cases. For a circular hole of radius $b$ in a thin wall $\psi = 8b^3/3$ and $\chi = 4b^3/3$ [4]. There are also analytical results for elliptic holes in a thin wall [5]. The polarizabilities for the case of a thick wall have been studied using a variational technique in [6] for circular holes, and in [7] for elliptic holes. There are also some approximate formulae for slots [8].

In the present paper, the polarizabilities of an annular cut in a thin wall are obtained and used to estimate the beam coupling impedances of the button-type beam position monitors (BPMs).

## II. General Analysis

When the wavelength is large compared to the hole size, the polarizabilities can be obtained from the electro- or magnetostatic problem: find the fields due to an aperture (hole) in a metal plane when it is illuminated from one side by a homogeneous static (normal electric or tangential magnetic) field.

### A. Integral Equations

Let a hole in the plane $z = 0$ be illuminated by a far magnetic field $H_0$ from $z > 0$ side. We assume that the hole center coincides with the origin of the plane coordinates $(u, v)$, and the field is directed along $\hat{u}$. One can decompose this far field as $H_0/2 + H_0/2 = H_0$ for $z > 0$, and as $H_0/2 - H_0/2 = 0$ for $z < 0$, and consider two separate problems — the symmetric and the antisymmetric one [6], [9]. For a zero thickness plane, the symmetric magnetic problem is trivial (the field is $H_0/2$ everywhere). The antisymmetric problem can be reduced to the integral equation [9] for the function $G(\vec{r}) = 2H_z(\vec{r}, 0)/H_0$

$$\int_h d\vec{r}' G(\vec{r}') K(\vec{r}, \vec{r}') = u , \tag{2}$$

where $\vec{r} = (u, v)$, the integration runs over the aperture, and the kernel is symmetric

$$K(\vec{r}, \vec{r}') = \frac{1}{4\pi^2} \int \frac{d\vec{\sigma}}{\sigma} e^{i\vec{\sigma}(\vec{r}-\vec{r}')} = \frac{1}{2\pi|\vec{r}-\vec{r}'|} . \tag{3}$$

If Eq. (2) is solved, the magnetic susceptibility is [9]

$$\psi_u = \int_h d\vec{r}' u G(\vec{r}') . \tag{4}$$

For an axisymmetric aperture, one can simplify Eq. (2) using $u = r \cos\varphi$, substituting $G(\vec{r}) = g(r) \cos\varphi$, and intergating over the polar angle $\varphi'$. It yields

$$\int_{[h]} dr' r' g(r') K_m(r, r') = r , \tag{5}$$

with the following kernel

$$\begin{aligned}
K_m(x, y) &= \int_0^\infty d\sigma J_1(\sigma x) J_1(\sigma y) \tag{6} \\
&= \theta(y-x) \frac{x}{2y^2} \,{}_2F_1\!\left(\frac{3}{2}, \frac{1}{2}; 2; \frac{x^2}{y^2}\right) + \{x \leftrightarrow y\} \\
&= \frac{xy}{2(x+y)^3} \,{}_2F_1\!\left(\frac{3}{2}, \frac{3}{2}; 3; \frac{4xy}{(x+y)^2}\right) ,
\end{aligned}$$

where $J_n(x)$ is the n-th order Bessel function of the first kind, and ${}_2F_1$ is the Gauss hypergeometric function. This kernel has a ln-singularity at $x = y$

$$\begin{aligned}
K_m(x, y) &\simeq \frac{8xy}{\pi(x+y)^3}\left(\ln\frac{x+y}{|x-y|} + 2\ln 2 - 2\right) \\
&\quad + O(|x-y|\ln|x-y|) . \tag{7}
\end{aligned}$$

The magnetic susceptibility in this case is

$$\psi = \pi \int_{[h]} dr\, r^2 g(r) \ . \tag{8}$$

In Eqs. (5) and (8) symbol $[h]$ denotes the interval of the radius-vector variation: $[h] = [0, b]$ for a circular hole of radius $b$, and $[h] = [a, b]$ for an annular cut with inner radius $a$ and outer radius $b$.

In a similar way, a solution $f(r)$ of the electrostatic problem satisfies the integral equation

$$\int_{[h]} dr'\, r' f(r') K_e(r, r') = 1 \ , \tag{9}$$

with a more singular $[O\left((x-y)^{-2}\right)]$ kernel

$$K_e(x, y) = \int_0^\infty d\sigma\, \sigma^2 J_0(\sigma x) J_0(\sigma y) \ . \tag{10}$$

The electric polarizability of the axisymmetric hole is

$$\chi = 2\pi \int_{[h]} dr\, r f(r) \ . \tag{11}$$

A solution $g(r)$ of the integral equation (5) must have the correct singular behavior near the thin metal edge: $g(r) \propto \Delta^{-1/2}$ when $\Delta = b - r \to 0$ or $\Delta = r - a \to 0$. For the problem (9), the function $f(r)$, which is proportional to the electric potential, must behave as $\sqrt{\Delta}$ near the edge to provide for the correct singularity $\Delta^{-1/2}$ of the electric field. In the case of a circular hole of radius $b$ the exact solutions of Eqs. (5) and (9) are known [4]. They are $g(r) = 4r/(\pi\sqrt{b^2 - r^2})$ and $f(r) = 2\sqrt{b^2 - r^2}/\pi$, substituting of which in (8) and (11) gives the polarizabilities of a circular hole cited in Introduction.

### B. Narrow Cut: Analytical Solution

Suppose the width $w = b - a$ of the gap is small, $w \ll b$. For a narrow annular cut, the electric polarizability can be approximated by that of a narrow (yet bented) slot of width $w$ and length $2\pi b \gg w$ as $\chi \simeq \pi^2 w^2 b/4$, see [8]. It is obvious that $\chi$ is small compared to $\psi$, since the normal electric field does not penetrate far through the narrow gap, unlike the tangential magnetic field on the parts of the annular cut which are parallel to its direction.

Introducing dimensionless variables $x = r'/b$ and $y = r/b$, we are looking for a solution of Eq. (5) in the form $g(x) = C(x)/\sqrt{(1-x)(x-\rho)}$, where $\rho = a/b$, and $C(x)$ is a regular function in the interval $[\rho, 1]$. For a narrow gap $\delta \equiv 1 - \rho \ll 1$, and one can expand $C(x)$ as $C(x) = C + O(\delta)$. Substituting this into Eq. (5) and keeping only the singular part (7) of the kernel (the rest would give corrections $O(\delta)$ to the RHS) leads to the equation

$$1 = \frac{C}{\pi} \int_\rho^1 \frac{dx\, [\ln(8/|x-y|) - 2]}{\sqrt{(1-x)(x-\rho)}} \ , \tag{12}$$

where we neglected terms $O(\delta \ln \delta)$ in the RHS. Replacing variables $x = 1 - u\delta$, $y = 1 - v\delta$, and using the identity

$$\int_0^1 \frac{du\, \ln|u - v|}{\sqrt{u(1-u)}} = -2\pi \ln 2 \ ,$$

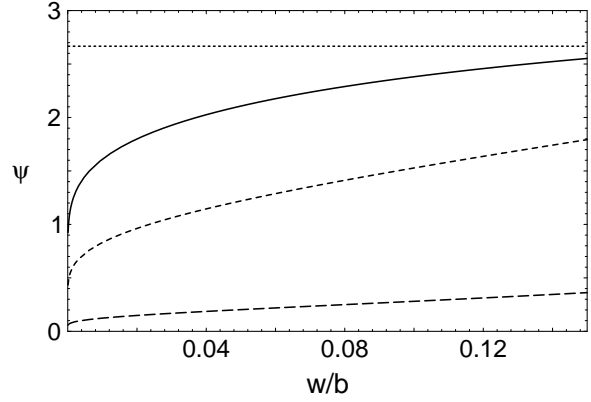

Figure. 1. Magnetic susceptibility (in units of $b^3$) of a narrow annular cut versus its relative width $w/b$: solid line for (14), long-dashed line for octagon model (15), and short-dashed line for slot model (16). The dotted line shows the susceptibility of the circular hole $\psi/b^3 = 8/3$.

we get from (12)

$$C = [\ln(32/\delta) - 2]^{-1} \ . \tag{13}$$

Then from Eq. (8) the magnetic susceptibility of a narrow ($w = b - a \ll b$) annular cut in a thin plate is

$$\psi = \frac{\pi^2 b^2 a}{\ln(32b/w) - 2} \ . \tag{14}$$

It is interesting to compare Eq. (14) with the estimate [8] obtained by approximating the annular cut with an octagon and using the magnetic susceptibilities for narrow slots:

$$\psi_o = \frac{4}{3} \left(\frac{\pi}{4}\right)^4 \frac{b^3}{\ln(2\pi b/w) - 7/3} \ . \tag{15}$$

While the behavior is similar, this estimate is a few times smaller than (14), see Fig. 1. Moreover, even a more extreme model — two long slots of length $2b$ and width $w$ oriented parallel to the magnetic field — give the susceptibility

$$\psi_m = \frac{4}{3} \frac{\pi b^3}{\ln(16b/w) - 7/3} \ , \tag{16}$$

which is still smaller than Eq. (14), see Fig. 1.

As seen from Fig. 1, the susceptibility (14) becomes close to that of a circular hole for relatively narrow gaps, $w/b \geq 0.1$. To check this surprising result, and to find the applicability range for Eq. (14), we proceed below with a variatonal study of Eq. (5).

### C. Wide Cut: Variational Approach

An elegant variational technique for polarizabilities has been developed in [6]. Multiplying Eq. (5) by $rg(r)$ and integrating over $r$, we convert it to the following variational form for the magnetic susceptibility $\psi$

$$\frac{\pi b^3}{\psi} = \frac{\int_\rho^1 x\,dx \int_\rho^1 y\,dy\, g(x) K_m(x, y) g(y)}{\left[\int_\rho^1 x^2\, dx\, g(x)\right]^2} \ . \tag{17}$$

A solution $g(x)$ of Eq. (5) minimizes the RHS of Eq. (17). We are looking for it in the form of a series

$$g(x) = \sum_{n=0}^{\infty} c_n g_n(x) \qquad \text{with} \qquad (18)$$

$$g_0(x) = \frac{1}{\sqrt{(1-x)(x-\rho)}}, \qquad g_k(x) = T_{k-1}\left(\frac{2x-\rho-1}{1-\rho}\right),$$

where $c_n$ are unknown coefficients, and $T_n(x)$ are Chebyshev's polynomials of the first kind. Denoting $d_n = \int_\rho^1 dx\, x^2 g_n(x)$ and $a_n = c_n d_n$, we define the matrix

$$K_{kn} = \int_\rho^1 x\,dx \int_\rho^1 y\,dy\, g_k(x) K_m(x,y) g_n(y)/(d_k d_n), \quad (19)$$

and convert Eq. (17) into the following form

$$\frac{\pi b^3}{\psi} = \frac{\sum_{k,n} a_k K_{kn} a_n}{\left(\sum_n a_n\right)^2}. \qquad (20)$$

Following [6], one can prove that minimizing the RHS of Eq. (20) yields

$$\psi = \pi b^3 \sum_{k,n} \left(K^{-1}\right)_{kn}, \qquad (21)$$

where matrix $K^{-1}$ is the inverse of the matrix $K$, Eq. (19). The further procedure is straightforward: $n$th iteration ($n = 0, 1, 2, \ldots$) corresponds to the matrix (19) truncated to the size $(n+1) \times (n+1)$. In the zeroth iteration the truncated matrix is merely a number, $K_{00}$; it corresponds to the analytical study of Sect. II.B. All integrations and matrix inversions have been carried out using *Mathematica*. Calculations show that only even terms of the series (18) contribute, i.e. $c_1 = c_3 = \ldots = 0$, and, effectively, one can use $g = g_0 + c_2 T_1 + c_4 T_3 + \ldots$, and squeeze matrix $K$ removing odd lines and rows. The results for $\psi$ versus the cut width are shown in Fig. 2. As one can see, the zeroth iteration, as well as the analytical solution (14), is good for narrow gaps, $w/b \leq 0.15$, but it is also not bad for wide ones. The process practically converges in three iterations (effective 0,1,2) for the whole range of the cut width $0 \leq w/b \leq 1$.

### III. Impedances

The beam-chamber coupling impedances can be obtained using formulas from [1], [2], [3] and polarizabilities found in Sect. II. For example, a narrow annular cut of radius $b$ and width $w \ll b$ on the thin wall of a circular pipe of radius $r \gg b$ produces the longitudinal impedance

$$Z(\omega) = -\frac{iZ_0\omega(\psi-\chi)}{8\pi^2 cr^2} \simeq -\frac{iZ_0\omega b^3}{8cr^2\left[\ln(32b/w) - 2\right]}. \qquad (22)$$

This result can be used to estimate the impedance of a button-type BPM. Taking into account the wall thickness reduces the estimate, cf. [6], [8]. For other cross sections of the chamber, $Re\, Z$, and the transverse impedance, see [3] and references therein. Note that the impedance (22) of a narrow cut with $w/b > 0.05$ is larger than (but less than twice) that of a circular hole with radius $b$, and tends to the last one when $w \to b$.

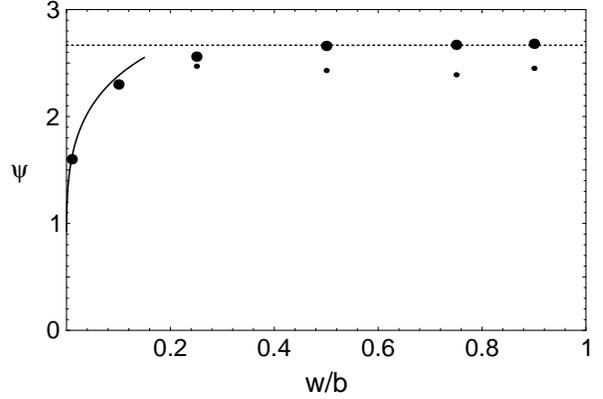

Figure. 2. Magnetic susceptibility (in units of $b^3$) of an annular cut versus its relative width $w/b$. Thin points show the zeroth iteration, and thick points are for the second one. The solid line corresponds to Eq. (14) for narrow gaps, and the dotted line is the limit of a circular hole, $w = b$.

### IV. Conclusions

The magnetic susceptibility of an annular cut in a thin wall is calculated using the analytical and variational methods. The estimate for the coupling impedance of a button-type BPM is obtained.

The electro- and magnetostatic problems considered above can also be solved numerically. With boundary conditions which ensure a given homogeneous field far from the aperture plane, a static electric or magnetic potential can be computed using standard codes. We have done this for the electric polarizability of an axisymmetric aperture, in which case the problem is effectively a 2-D one, using the *POISSON* code. Results for a narrow gap and a hole coincide with the expected ones. The case of a thick wall can be also studied in this way. Unfortunately, for the magnetic problem, as well as for an arbitrary aperture, this approach requires 3-D codes and cumbersome computations.

We plan a further study using different methods to take into account the effects due to the wall thickness.

The author would like to thank Dr. R.L. Gluckstern and Dr. R.K. Cooper for useful discussions.